\documentclass[aps,twocolumn,prd,superscriptaddress,floatfix]{revtex4-1}

\usepackage[utf8]{inputenc}

\usepackage{amsmath}
\usepackage{mathtools}
\usepackage{amsfonts}
\usepackage{mathrsfs}
\usepackage{bbm}
\usepackage{dsfont}
\usepackage{nicefrac}

\usepackage{graphicx}
\usepackage{xcolor}

\usepackage{placeins}

\usepackage{xspace}
\usepackage{siunitx}

\usepackage{url}
\usepackage[colorlinks=true,urlcolor=blue,linkcolor=blue,citecolor=blue,hypertexnames]{hyperref}
\usepackage[nameinlink]{cleveref}
\usepackage{braket}
\usepackage{simplewick}
\usepackage{float}

\usepackage{xifthen}
\usepackage{xcolor}
\hypersetup{
	colorlinks,
	linkcolor={red!75!black},
	citecolor={blue!75!black},
	urlcolor={blue!75!black}
}

\usepackage{csquotes}
\usepackage[shortcuts]{extdash}	
\usepackage{comment}
\usepackage{xifthen}

\setkeys{Gin}{width=0.48\textwidth}
\def\App#1{App.~\ref{#1}}

\graphicspath{{./figures/}}

\renewcommand{\vec}[1]{\mathbf{#1}}

\newcommand{\imag}{\mathrm{i}}
\newcommand{\contourC}{\mathcal{C}}

\DeclareMathOperator{\diff}{d\!}

\DeclareMathOperator{\deltac}{\delta_\mathcal{C}}

\DeclareMathOperator{\defeq}{\vcentcolon=}

\DeclareMathOperator{\Tr}{Tr\!}
\DeclareMathOperator{\T}{{\cal T}}
\DeclareMathOperator{\Tc}{{\T}_{\!\!\contourC}}

\DeclareMathOperator{\phibar}{\hspace{.1em}\bar{\phi}\hspace{.1em}}

\newcommand*{\sspace}{\hspace{.1em}}
\newcommand*{\nsspace}{\hspace{-.1em}}

\def\eq#1{(\ref{#1})}

\def\eqref#1{(\ref{#1})}

\DeclareSymbolFont{bbold}{U}{bbold}{m}{n} 
\DeclareMathSymbol{\one}{\mathalpha}{bbold}{"31}	
\newcommand*{\oneO}{\one}

\newcommand{\gettitle}{Causal Temporal Renormalisation Group Flow of the Energy-Momentum Tensor}

\hypersetup{
	pdftitle={\gettitle},
	pdfauthor={Heller, Pawlowski},
	pdfkeywords={time evolution} {energy-momentum tensor} {trace anomaly},
	bookmarksopen=true,
	bookmarksopenlevel=2,
	bookmarksnumbered=true
}

\begin{document}

\title{\gettitle}

\author{Markus Heller}
\affiliation{Institut f\"ur Theoretische Physik,
	Universit\"at Heidelberg, Philosophenweg 16,
	69120 Heidelberg, Germany
} 

\author{Jan M. Pawlowski}
\affiliation{Institut f\"ur Theoretische Physik,
	Universit\"at Heidelberg, Philosophenweg 16,
	69120 Heidelberg, Germany
}
\affiliation{ExtreMe Matter Institute EMMI,
	GSI, Planckstr. 1,
	64291 Darmstadt, Germany
}

\begin{abstract}
We derive the temporal renormalisation group flow of the energy-momentum tensor at the example of a general scalar theory. The local causal structure of the temporal renormalisation group flow allows to monitor and control causality, unitarity and general conservation laws at each infinitesimal renormalisation group step. We explore energy-conserving truncations in a comparison of generic flows and the causal temporal flow of the energy-momentum tensor. We also observe that the temporal regulator preserves scale invariance, which is violated for generic momentum  regulators. Specifically we discuss the relation of these terms to the trace anomaly of the energy-momentum tensor. Moreover, we show that the causal temporal flow of the energy-momentum tensor can be integrated analytically, and demonstrate, that the result is consistent with energy conservation.
\end{abstract}

\maketitle

\section{Introduction}
\label{sec:intro}

Understanding the time evolution of quantum field theories in and out of equilibrium is relevant for a variety of different physical systems ranging from inflationary cosmology over the dynamics of phase transitions in the standard model and heavy ion collisions to table top experiments with ultracold atoms.

The framework of the temporal functional renormalisation group (t-fRG) constitutes a manifestly causal approach that gives access to the dynamics of correlation functions in quantum field theories \cite{Gasenzer:2007za, Gasenzer:2010rq, Corell:2019jxh}. As opposed to standard flows in momentum space with imaginary time, the t-fRG flow is a real-time flow. Causality is ensured by a causal temporal regulator, that suppresses all quantum fluctuations beyond the cutoff time $\tau$, hence terminating the Keldysh time contour at the cutoff time. 

In the present work we discuss the time evolution of the energy-momentum tensor (EMT). The EMT carries some of the most important conservation laws of a quantum field theory at hand, and its correlations are relevant for the construction and benchmarking of kinetic theory, hydrodynamics and transport models. Via its trace it also gives access to the scale properties of the theory including the quantum scale or trace anomaly. In standard functional renormalisation group (fRG) applications, the regularisation in momentum space explicitly breaks scale invariance. Interestingly, we observe that this is not the case for the causal temporal regulator of the t-fRG, and we will discuss the underlying reasons in detail. 

Moreover, in non-equilibrium applications the flows of the EMT and its correlations serve as benchmark observables for approximation schemes: they carry important conservation laws whose conservation can be controlled in each flow step. 

Finally, in \cite{Corell:2019jxh} it has been shown that due to the locality and causality of the t-fRG flows, they can be integrated analytically. This leads to novel one-loop exact  diagrammatic representations of correlation functions. Here, we will study the integrated flow of the EMT in view of energy conservation within given approximation schemes. 

The present work starts with a brief review of the relevant aspects of the t-fRG framework in \Cref{sec:t-fRG}. In \Cref{sec:flowEMT} we derive the general and the causal temporal flow of the EMT and we also discuss the flow of the  trace anomaly. The integrated flow of the EMT is derived and discussed in \Cref{sec:intflow}. In \Cref{sec:consistency}, we perform a non-trivial consistency check of the integrated flow: it carries the conservation of energy and momentum in the quantum theory at hand. \Cref{sec:conclusion} contain brief conclusions.

\section{The Temporal Functional Renormalisation Group}
\label{sec:t-fRG}

\begin{figure}[t]
	\centering
\includegraphics[width=.41\textwidth]{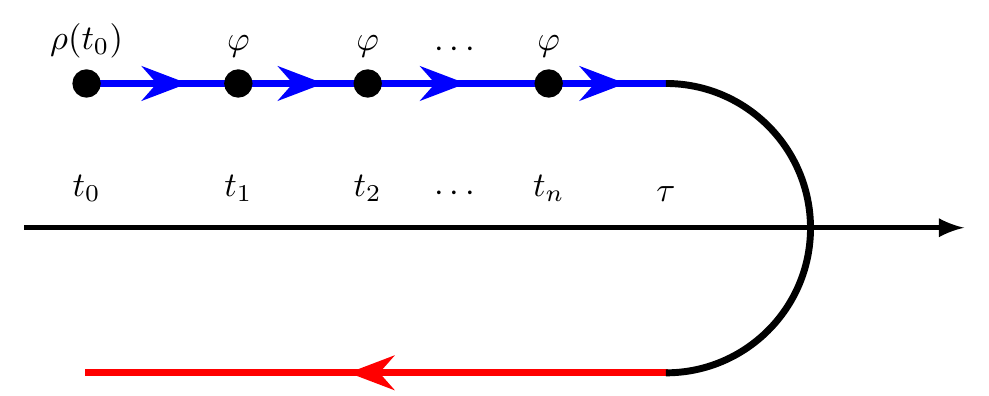}
\caption{Illustration of the causality-property \cref{eq:ctpsmaller}: An $n$\-/point function $\Gamma^{(n)}_\tau$ derived from \cref{eq:gammatau}, with all $t_1,\dots,t_n<\tau$, equals the fully dressed $n$\-/point function. Here, the upper (blue) branch, referred to as $\contourC^+$, encodes time-ordered time evolution, and the lower (red) branch, referred to as $\contourC^-$, encodes anti--time-ordered evolution. The density matrix at the initial time $t_0$ is given by $\rho(t_0)$, and $\varphi$ is the field operator in the Schr\"odinger picture.}
	\label{fig:ctpdiagsmaller}
\end{figure}

Here we briefly review important aspects of the t-fRG approach, more details can be found in \cite{Gasenzer:2007za, Gasenzer:2010rq, Corell:2019jxh}. Its central object is the effective action $\Gamma[\phi]$, defined on the Schwinger-Keldysh closed time path (CTP) $\contourC$ , \cite{Schwinger:1960qe, Keldysh:1964ud}. For recent introductions see e.g.\ \cite{Calzetta:2008iqa, Stefanucci2010, Berges:2015kfa}, for momentum cutoff flows on the CTP see e.g.\ \cite{Berges:2008sr, Huelsmann:2020xcy, Tan:2021zid} and the review \cite{Dupuis:2020fhh}. The effective action $\Gamma[\phi]$  is the generating functional of the one-particle irreducible parts of correlation functions, and can be derived from the generating functional of full correlation functions,   
\begin{align}
	Z[J;{\rho}] = \Tr\bigg[{\rho}(t_0)\Tc\exp \bigg\lbrace \imag
	\int\displaylimits_{\mathrlap{\contourC(x)}}
	J(x)\,\varphi(x) \bigg\rbrace \bigg] \,,
\label{eq:Z}
\end{align}
with the time ordering $\Tc$ of the Keldysh contour, 
\begin{align}
	\int\displaylimits_{\contourC(x)} \defeq 
	\Bigg[\, \int_{t_0, \contourC^+}^{\infty} \diff x^0 \;
	- \int_{t_0,\contourC^-}^{\infty} \diff x^0 \,\Bigg]\int_{\vec{x}}\;.
\end{align}
Here $x$ denotes a $D=d+1$ dimensional vector with time component $x^0$ and $d$-dimensional spatial component $\bf x$. Naturally, the field operator $\varphi$ lives on the Keldysh contour, depicted in \Cref{fig:ctpdiagsmaller} and $\rho(t_0)$ is the density matrix at the initial time $t_0$. We now introduce a temporal cutoff by terminating the Keldysh contour at a final time $\tau$. Evidently, this does not affect correlation functions of fields, where all times are smaller than $\tau$, and hence the evolution of the theory with $\tau$ is the genuine time evolution of the system. Such a cutoff is implemented by 
\begin{align}
	Z_{\tau}[J] 
	&= \exp\left\{-\frac{\imag}{2} \int\displaylimits_{\contourC(x),\contourC(y)}
	\frac{\delta}{\delta J(x)} R_{\tau}(x,y) \frac{\delta}{\delta J(y)} \right\} Z[J]\,,
	\label{eq:regfunc}
\end{align}
with the temporal causal regulator $R_\tau$, 
\begin{align}
	-\imag R_{\tau}(x,y) = \left\lbrace
	\begin{aligned}
		&\infty \quad &&x^0=y^0 > \tau, \vec{x} = \vec{y} \\
		& 0 \quad &&\mathrm{otherwise}
	\end{aligned}
	\right. \,.
	\label{eq:reg}
\end{align}
Evidently, the generating functional vanishes for currents $J$ with support at times larger than $\tau$.  
The causal dynamics of the 1PI correlation functions is encoded by that of the effective action $\Gamma_{\tau}[\phi]$, the Legendre transform of 
the Schwinger functional $	W_{\tau}[J]= \log Z_\tau[J]$,  
\begin{align}
	\Gamma_{\tau}[\phi] 
	\,\defeq\, 
	W_{\tau}[J] 
	\,-\, 
	J\cdot\phi
	\,-\frac{1}{2}\, 
	\phi\cdot R_{\tau}\cdot\phi\,,
	\label{eq:gammatau}
\end{align}
with the short hand notation ${f\cdot g=\int\displaylimits_{\contourC(x)} f(x)g(x)}$ for the CTP integral. Moeover, $J=J[\phi]$ is the solution of $\delta W_\tau/\delta J = \phi$. With the sharp causal regulator \eq{eq:reg}, all fluctuations beyond the cutoff time $\tau$ are completely suppressed. Hence, $\Gamma_{\tau}[\phi]$ contains all fluctuations up to times $\tau$, and the full dynamics of the quantum theory is obtained in the limit $\tau\rightarrow\infty$. An infinitesimal time step is 
governed by the  temporal flow equation, see \cite{Gasenzer:2007za, Gasenzer:2010rq, Corell:2019jxh}, 
\begin{align}
	\partial_{\tau}\Gamma_{\tau}[\phi] = \frac{1}{2} 
	\Tr \Big[G_{\tau}[\phi]\cdot\partial_{\tau}R_{\tau}\Big]\,,
	\label{eq:floweq}
\end{align}
with the full field-dependent propagator $G_\tau[\phi]$, 
\begin{align}
	G_{\tau}[\phi](x,y) 
	=&\, \left[ \frac{\imag }{\Gamma^{(2)}_{\tau}[\phi]+R_{\tau}}\right] \! (x,y)\,. 
	\label{eq:gphi}
\end{align}
The propagator in \cref{eq:gphi} vanishes identically for times $x_0>\tau$ or $y_0>\tau$ due to the regulator. In \cref{eq:gphi} we used the notation $\Gamma^{(n)}$ for $nth$ derivatives of the effective action w.r.t.\ $\phi$. We emphasise that the mathematically sound definition of the flow \cref{eq:floweq} requires the careful evaluation of the regularisation limit of the product of singular distributions  $G_\tau$ and $\partial_{\tau}R_{\tau}$. This has been discussed in detail in \cite{Gasenzer:2007za, Gasenzer:2010rq} and in particular in \cite{Corell:2019jxh, HellerDiss}. It leads to a technically straightforward implementation in terms of a modified product of distributions, which we call the $*$-product, see \Cref{app:starproduct}. This facilitates in particular the derivation of analytically integrated flows considered later in \Cref{sec:intflow}. 

\begin{figure}[t]
	\centering
	\includegraphics[width=.41\textwidth]{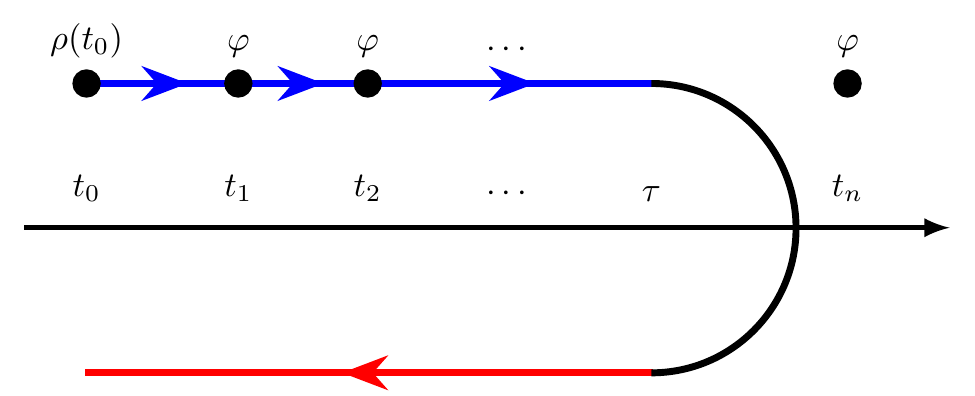}
	\caption{Illustration of the causality-property \cref{eq:ctplarger}: An $n$\-/point function $\Gamma^{(n)}_\tau$ derived from \cref{eq:gammatau}, with at least one time $t_n$ larger than $\tau$, remains at its initial value. $\varphi$ at $t_n$ can not be placed on the contour, which is best discussed in terms of the sources in \eq{eq:regfunc}. A derivative with respect to $J(t_n>\tau)$ leads to correlation functions involving $\varphi(t_n)$ with vanishing measure due to the causal regulator \cref{eq:reg}. Hence, the 1PI correlation functions remain at their initial values.}
	\label{fig:ctpdiaglarger}
\end{figure}
The unique causality properties of the t-fRG are best illustrated at the example of the $n$-point correlation functions $\Gamma^{(n)}_{\tau}(x_1,\dots,x_n)$. For times $t_1\leq t_2 \leq \cdots \leq t_n \leq \tau$ with $t_i=x^0_i$, the $\tau$-dependent correlation function agrees with the full correlation function, 
\begin{align}
	\Gamma^{(n)}_{\tau}(x_1,\dots,x_n)= \Gamma^{(n)}(x_1,\dots,x_n)\,, 
	\label{eq:ctpsmaller}
\end{align}
where $\Gamma^{(n)}(x_1,\dots,x_n):= \Gamma^{(n)}_{\tau=\infty}(x_1,\dots,x_n)$. This situation is depicted in \Cref{fig:ctpdiagsmaller}. Note that in the absence of field insertions, the contributions from the time evolution operators on the upper part $\contourC^+$ of the Keldysh contour are cancelled by the contributions from the lower part $\contourC^-$. For this reason the extension of the Keldysh contour beyond $x^0_n$ does not matter, and the flow for $\tau> t_n$ does not change the respective $\Gamma^{(n)}_\tau$ with $t_n<\tau$. 

In turn, if at least one time is larger than the cutoff time, that is $t_n>\tau$, we have, 
\begin{align}
	\Gamma^{(n)}_{\tau}(x_1,\dots,x_n) = \Gamma^{(n)}_{t_0}(x_1,\dots,x_n)\,.  
	\label{eq:ctplarger}
\end{align}
\Cref{eq:ctplarger} originates from the fact, that the causal regulator $R_\tau$ suppresses \emph{all} fluctuations for times later than $\tau$. This situation is depicted in \Cref{fig:ctpdiaglarger}. The fact that the field $\varphi$ at $t_n$ can not be placed on the contour, is best interpreted in terms of the sources in \cref{eq:regfunc}: sources with times larger than $\tau$ are completely suppressed by the causal regulator. Hence, a derivative with respect to such a source vanishes. In turn, the 1PI correlation functions simply keep their initial values. 

Let us emphasise that the properties \cref{eq:ctpsmaller} and \cref{eq:ctplarger} are derived from the flow \cref{eq:floweq}. The only additional assumption needed is that the classical vertices are diagonal in time which encodes the locality of the microscopic interactions. 

The causality properties \cref{eq:ctpsmaller} and \cref{eq:ctplarger} lead to an important identity unique to the present causal t-fRG-approach, 
\begin{align}
	\partial_\tau G_\tau[\phibar]=\imag\,(G_\tau\partial_\tau R_\tau G_\tau)[\phibar]\,, 
	\label{eq:gtau}
\end{align}
for a detailed discussion see \cite{Corell:2019jxh}. Here, $\phibar$ is the physical, in general space-time dependent, background which is given by $\phi^+=\phibar=\phi^-$ where $\phi^\pm$ are the fields on $\contourC^\pm$. Notably, the term involving $\partial_\tau\Gamma^{(2)}_\tau[\phibar]$ is absent for the causal temporal flow. This remarkable fact is deeply rooted in the locality and causality of the employed regulator, and is also linked to functional optimisation of the fRG \cite{Pawlowski:2005xe}. The above properties of local causal flows also imply 
\begin{align}\label{eq:Gtau}
	G_\tau[\bar{\phi}](x,y)=G_{\tau=\infty}[\bar{\phi}](x,y)\theta(\tau-x^0)\theta(\tau-y^0)\,. 
\end{align}
\Cref{eq:Gtau} entails that $G_\tau[\bar{\phi}]$ is either the fully dressed propagator or it vanishes. This concludes our brief overview on the t-fRG approach.

\section{The Flow of the Energy-Momentum Tensor}
\label{sec:flowEMT}

In the present Section we derive the flow of the energy-momentum tensor (EMT), following the derivation in \cite{Pawlowski2021prep}. We consider a real scalar theory with microscopic three- and four-point interactions, but the derivations trivially generalise to generic quantum field theories. The classical action of the real scalar theory is given by
\begin{align}\nonumber
	S[\varphi]=\int_{\contourC}\bigg\lbrace&\frac{1}{2}\,
	\partial^\rho\varphi(x)\partial_\rho\varphi(x)-
	\frac{m^2}{2}\,\varphi(x)^2\\[1em]
	-&\frac{\lambda_3}{3!}\,\varphi(x)^3
	-\frac{\lambda_4}{4!}\,\varphi(x)^4\bigg\rbrace\,.
	\label{eq:action}
\end{align}
For the derivation of the EMT we augment \cref{eq:action} with a background metric $g_{\mu\nu}$ with  $\int_{\contourC}\rightarrow\int_{\contourC}\sqrt{-g(x)}$. The EMT is the  metric derivative of the action, evaluated at the flat Minkowski background, $g_{\mu\nu}=\eta_{\mu\nu}$, 
\begin{align}
	T_{\mu\nu}(x)=
	 \left. \frac{2}{\sqrt{-g(x)}}
	\frac{\delta }{\delta g^{\mu\nu}(x)}S[\varphi,g]
	\right|_{g^{\mu\nu}=\eta^{\mu\nu}}\,.
	\label{eq:EMT}
\end{align}
Here, $g=\det g_{\mu\nu}$ is the determinant of the metric. For the action \cref{eq:action} we are  led to, 
\begin{align}\nonumber
	T_{\mu\nu}(x) =&\,  \partial_\mu\varphi\sspace\partial_\nu\varphi \\[1ex] 
&	\hspace{-.8cm}- \eta_{\mu\nu}\Big[\frac{1}{2}\partial^\rho\nsspace\varphi\sspace\partial_\rho\varphi	
	-\frac{m^2}{2}\varphi^2-\frac{\lambda_3}{3!}\varphi^3 - \frac{\lambda_4}{4!}\varphi^4\bigg]\;.
	\label{eq:EMTscalar}
\end{align}
For some details on metric-derivatives see \App{app:identities}. 

\subsection{The General Flow of the EMT}
\label{sec:generalflow}

The definition \cref{eq:EMT} straightforwardly generalises to the regulated effective action 
$\Gamma_\tau$, and the quantum EMT is given by 
\begin{align}
	\mathcal{T}_{\mu\nu,\tau}[\phi](x)=
	 \left. \frac{2}{\sqrt{-g(x)}}
	\frac{\delta }{\delta g^{\mu\nu}(x)}\Gamma_\tau[\phi,g]
	\right|_{g^{\mu\nu}=\eta^{\mu\nu}}\,, 
	\label{eq:vevEMT}
\end{align}
which is related to the expectation value of the classical EMT, $\langle T_{\mu\nu} \rangle$. Importantly,  additional terms stem from the renormalisation of the path integral and provide the scale anomaly. Following \cite{Pawlowski2021prep}, we derive the flow of the EMT as the $\tau$-derivative of \cref{eq:vevEMT}, 
\begin{align}
	 \partial_\tau \mathcal{T}_{\mu\nu,\tau}(x)=
	 \frac{1}{\sqrt{-g(x)}}
	\frac{\delta }{\delta g^{\mu\nu}(x)}
	\Bigl(\Tr\, G_\tau\cdot\dot{R}_\tau \Bigr)\,,
	\label{eq:metricderivativegammatau}
\end{align}
with $\partial_\tau R_\tau=\dot{R}_\tau$ and $g_{\mu\nu}\to\eta_{\mu\nu}$. For deriving  \cref{eq:metricderivativegammatau} we have commuted metric and $\tau$-derivatives. Since the flow of $\Gamma_\tau$ is local and causal, the flow of the EMT inherits these important properties. The trace in \cref{eq:metricderivativegammatau} is performed with basis functions that are normalised with $1/\sqrt{-g}$, providing a covariant basis. 

We proceed with computing the metric-derivative in \cref{eq:metricderivativegammatau}. The derivative of the trace can be evaluated according to \cref{eq:intderivative} and more details can be found in \App{app:identities}. For the propagator we find
\begin{align}\nonumber 
	\frac{\delta G_{\tau,ab}}{\delta g^{\mu\nu}(x)} 
	=&\,
	\imag \left[G_{\tau}\cdot
	\,\frac{\delta( \Gamma^{(2)}_\tau + R_\tau )}{\delta g^{\mu\nu}(x)}\cdot\,G_\tau \right]_{ab}\\[1ex]
	&+\frac12\,g_{\mu\nu}(x)\big[ G_{\tau,ax} \oneO_{bx}+G_{\tau,bx} \oneO_{ax}\big]\,.
	\label{eq:Gderivative}
\end{align}
Here, we use a shorthand notation denoting space-time arguments as indices. The first term on the right-hand side originates from its inverse relation to $\Gamma^{(2)}_\tau$, cf.~\cref{eq:gphi}. The second term comes from the covariant definition of the inverse in line with the present covariant basis, leading to $G\cdot G^{-1}=\oneO$, and $\oneO$ is defined in \cref{eq:one}. Note that the operator product on the left-hand side also contains a factor of $\sqrt{-g}$. These appearances of the metric are hit by the metric-derivative and produce the second term in \cref{eq:Gderivative}. Notably, this term cancels with the metric derivative of the trace in \cref{eq:metricderivativegammatau}.
Finally, we arrive at 
\begin{align}\nonumber
	&\partial_\tau \mathcal{T}_{\mu\nu,\tau}(x)
	=\Tr\Bigg[G_\tau \cdot
	  \frac{1}{\sqrt{-g(x)}}
	\frac{\delta \dot{R}_\tau}{\delta g^{\mu\nu}(x)}\Bigg]\\[1ex]
	&+\imag\Tr\Bigg[\big(G_\tau\cdot\dot{R}_\tau\cdot G_\tau\sspace\big) \cdot
	 \frac{1}{\sqrt{-g(x)}}
	\frac{\delta \big( \Gamma^{(2)}_\tau + R_\tau \big)}{\delta g^{\mu\nu}(x)}\Bigg]\,,
	\label{eq:genflowEMT}
	\raisetag{-1ex}
\end{align}
with $g^{\mu\nu}\rightarrow\eta^{\mu\nu}$. We emphasise that the above derivation holds true for general regulators, and in particular for standard momentum space regulators, see \cite{Pawlowski2021prep}.

\subsection{Causal Temporal Flow of the EMT}
\label{sec:causalflow}

Next, we turn to the causal structure of the temporal flow. We shall see that this causal structure leads us to a novel, unique result: The regulator terms in \cref{eq:genflowEMT} drop out completely in contradistinction to momentum space regulators used in \cite{Pawlowski2021prep}. This is due to the fact that the causal regulator, as opposed all other regulators, does not introduce violations of scale invariance. This fact and its implications with regard to the trace anomaly of the EMT are discussed in more detail in \Cref{sec:discussion}.

For proving the absence of regulator terms in the causal temporal flow of the EMT we first use \cref{eq:gtau}, only valid for the present sharp temporal regulator. This identity allows us to combine the regulator terms in \cref{eq:genflowEMT} into a total $\tau$-derivative, 
\begin{align}\nonumber 
	G_\tau\cdot\frac{1}{\sqrt{-g(x)}}\frac{\delta \dot{R}_\tau}{\delta g^{\mu\nu}(x)}
	+\dot{G}_\tau \cdot \frac{1}{\sqrt{-g(x)}}\frac{\delta R_\tau}{\delta g^{\mu\nu}(x)}\\[1em]
	=\frac{1}{2}g_{\mu\nu}(x)\,\partial_\tau\big(G_\tau\cdot R_\tau\big)\,,
\label{eq:TotalDert}\end{align}
where we have suppressed the field dependences in $G_\tau$. In the first line of \cref{eq:TotalDert} we have used  \eq{eq:metricderivativereg}: both, $R_\tau$ and $\partial_{\tau} R_\tau$ are proportional to the covariant unity, $\oneO$ defined in \cref{eq:one}, and hence their metric derivative is proportional to $R_\tau$ and $\partial_{\tau} R_\tau$ respectively with the same (metric) prefactors. Now we use that 
\begin{align}
	 \Bigl(R_{\tau}\cdot G_{\tau}\Bigr)[\phibar](x,y)=\left\{
	\begin{aligned}
		0\phantom{\deltac_{\!,xy}} \quad \text{if}\, x^0 \leq \tau \\
		\nsspace\nsspace\phantom{0}\deltac_{\!,xy} 
		\sspace\sspace\quad\text{if}\, x^0 > \tau
	\end{aligned}
	\right.\,.
	\label{eq:rtimesg}
\end{align}
where we have assumed $x^0>y^0$. This follows straightforwardly from the local causal properties of the temporal regulator \cref{eq:reg}, see \cite{HellerDiss} for more details. Its $\tau$-derivative and the trace leaves us with a constant theory-independent term which we simply drop. 

Then, we arrive at the remarkable fact that the metric-derivatives of the causal regulator do not contribute to the temporal flow of the EMT, which now reads
\begin{align}
	\hspace{-.05cm}\partial_\tau \mathcal{T}_{\mu\nu,\tau}[\phibar](x)
	=\frac{\imag}{2}\Bigl[ \Tr\, (G_\tau\cdot\dot{R}_\tau\cdot G_\tau)\cdot \mathcal{T}^{(2)}_{\mu\nu,\tau} \Bigr][\phibar](x), 
	\label{eq:causalflowEMT}
\end{align}
with 
\begin{align}
	\mathcal{T}^{(2)}_{\mu\nu,\tau}[\phi](x,y,z)	= \frac{2}{\sqrt{-g(x)}}
	\frac{\delta \Gamma^{(2)}_\tau[\phi](y,z) }{\delta g^{\mu\nu}(x)}\,, 
\label{eq:T2}\end{align}
\Cref{eq:causalflowEMT} has the standard form of the flow of composite operators derived in \cite{Pawlowski:2005xe}, see also \cite{Pagani:2016pad} and the review \cite{Dupuis:2020fhh}. It does not hold for momentum space regulators that introduce a (further) breaking of scale symmetry, see also \cite{Morris:2018zgy}. 

Using again the key identity \cref{eq:gtau}, only valid for the present local causal regulator, we are led to 
\begin{align}
	\partial_\tau \mathcal{T}_{\mu\nu,\tau}[\phibar](x)
	=\frac{1}{2}\Bigl[ \Tr\, \partial_\tau G_\tau \cdot \mathcal{T}^{(2)}_{\mu\nu,\tau} \Bigr][\phibar](x), 
	\label{eq:causalflowEMTFinal}
\end{align}
our final concise form of the EMT-flow. \Cref{eq:causalflowEMTFinal} is also straightforwardly generalised to EMT-correlations.  

\subsection{Properties of the causal temporal EMT-flow}
\label{sec:discussion}

As already mentioned above, \cref{eq:causalflowEMT} is the standard flow equation for composite operators. This equation does not hold true for the flow of the EMT for general regulators, given by  \cref{eq:genflowEMT}. Indeed, for momentum space regulators the terms in \cref{eq:genflowEMT} that originate from the metric dependence of the regulator generate (part of) the scale or trace anomaly. This can be understood as follows: For an infrared momentum space cutoff the theory tends towards the classical one in the ultraviolet as all momentum fluctuations are suppressed. For classical scale invariant theory, the trace of the classical EMT $T^\textrm{cl}$ vanishes as does $T^{\textrm{cl},(2)}{}_{\mu}{}^{\mu}$. If the full EMT at the initial time $t_0$ is identified with the classical EMT, $\mathcal{T}_{\mu\nu,t_0}[\phibar](x) =T_{\mu\nu}^{\textrm{cl}}[\phibar](x)$, the flow in \cref{eq:causalflowEMT} vanishes identically. However, the quantisation procedure typically breaks scale invariance due to the necessity of renormalisation, leading to a quantum scale anomaly. For an infrared regulator the quantisation of momentum modes and the respective scale anomaly is carried by the regulator and its metric dependence. Thus the flow successively generates the part of the scale anomaly generated by the quantum fluctuations of the respective momentum shells. In turn, the part of the scale anomaly carried by ultraviolet momentum modes larger than the initial infrared cutoff scale has to be present in the initial condition. 

The causal temporal flow of this work does not implement a momentum cutoff and the initial condition is the full renormalised quantum theory at hand at the initial time $t_0$. Specifically, the unitary quantum dynamics described by the causal temporal flow does not alter the UV properties of the theory. Thus, if we use renormalised initial conditions, the flow of the EMT is given by \cref{eq:causalflowEMT}. This will be discussed in more details in \cite{CausalFlowRen}, see also \cite{Corell:2019jxh, CorellDiss, HellerDiss}. We also remark that for some applications we may augment the present temporal cutoff with an additional spatial momentum cutoff. Then, the spatial momentum regularisation leads to a combined momentum and temporal flow of the EMT of the form \cref{eq:genflowEMT}. 

Next, we comment briefly on the use of the flow of the EMT \cref{eq:causalflowEMT} for monitoring or guaranteeing energy conservation in generic t-fRG truncations. For instance, we can use the flow of the EMT to close a truncation: while part of the flow of correlation functions present in a given truncation is derived from the flow of the effective action, the flow of the remaining set of correlation functions is derived from the flow of the EMT. This guarantees the unitarity of the flow of the EMT. Complementarily, the (integrated) flow of the EMT can be used numerically to correct for potential violations of energy conservation. 

We also remark that a suitable generalisation of the procedure as in  \cite{Garbrecht:2015cla} offers an attractive option. The general idea would be to alleviate potential violations of energy conservation caused by the truncation by using a suitably constructed background $\bar{\phi}$. Finally, the flow of the EMT in \cref{eq:causalflowEMT} can be used to simply monitor the conservation laws in supposedly unitary approximations.

\section{Integrated Causal Flow of the Energy-Momentum Tensor}
\label{sec:intflow}

Here, we discuss the analytic integration of the causal temporal flow of the EMT. This analytic integration is only possible due to the local and causal structure of the t-fRG approach leading to the simple flow \cref{eq:causalflowEMTFinal}. Due to the causality-properties discussed in \Cref{sec:t-fRG}, the only non-vanishing contributions to the integrated flow appear if the external time $x^0$ is equal to $\tau$. This results in 
\begin{align} \int_{t_0}^{\infty} \diff\tau\;
	\mathcal{T}^{(2)}_{\mu\nu,\tau}[\phibar](x)\,  
	\cdot\,\partial_\tau G_{\tau}[\phibar]=\mathcal{T}^{(2)}_{\mu\nu}[\phibar](x) \,*\, G[\phibar]\,,   
	\label{eq:shorthand}
\end{align}
with the $*$-product detailed in \Cref{app:starproduct}. It accounts for the well-defined product of causal singular distributions, see also \cite{Corell:2019jxh, HellerDiss}. Note also that the causality-properties \cref{eq:ctpsmaller} and \cref{eq:ctplarger} lead to causal constraints on the temporal flows of general operators. For instance, the causal constraint on the flow of the EMT is given by 
\begin{align}
	\partial_\tau \mathcal{T}_{\mu\nu,\tau}[\phibar](x) \propto \delta(\tau-x^0)*\theta(\tau-x^0)\,.
	\label{eq:causalconstraintEMT}
\end{align}
Importantly, the flow is only non-vanishing for $\tau=x^0$. This entails that the $\tau$-integral in \cref{eq:shorthand} terminates at $\tau = x^0+\xi$ with $\xi\to 0$, as the only contribution to the causal temporal flow of the EMT is proportional to $\delta(\tau-x^0)*\theta(\tau-x^0)$.

These types of constraints lead to the remarkable fact that the causal temporal flow can be integrated analytically. For the correlation functions themselves, this led to a novel one-loop exact functional relation for quantum field theories, see \cite{Corell:2019jxh, CorellDiss, HellerDiss}. Here we also arrive at a one-loop ecact relation for the integrated flow of the EMT, in comparison to the three-loop exact Dyson-Schwinger equation (DSE) for the EMT. The latter property originates in the fact that the EMT is a composite operator with up to forth order powers in the field, as we shall see later in \Cref{sec:consistency}. 
\begin{figure}[t]
		\includegraphics[width=.48\textwidth]{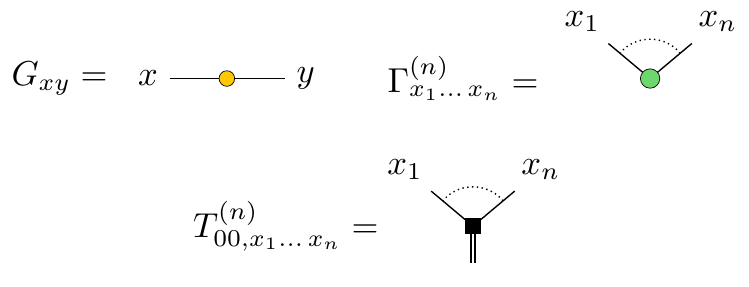}
		\caption{Symbols used in the graphical representation of the diagrams contributing to $\langle T_{00}[\varphi](x)\rangle$, cf.~\cref{eq:hamdiags} and \Cref{fig:hamdiags}. The black line with the orange circle represents the fully dressed field-dependent propagator $G_{xy}[\phi]=\imag\big[\Gamma^{(2)}[\phi]\big]_{xy}^{-1}$. The green circle with $n$ lines attached to it denotes the fully dressed $n$-point vertex $\Gamma^{(n)}_{x_1\dots\,x_n}[\phi]$. The black square with $n$ lines attached to it denotes the $n$th $\phi$\-/derivative of the classical energy $T^{(n)}_{00}[\phi](x,x_1\dots\,x_n)$, where the double line indicates the dependence on the external argument $x$.}
		\label{fig:legend}
\end{figure}

With the $*$\-/product, see \Cref{app:starproduct}, we express the integrated flow of the EMT as
\begin{align}
	\mathcal{T}_{\mu\nu}(x) - \mathcal{T}_{\mu\nu,\mathrm{init}}(x)
	=\frac{1}{2}\Tr\Big[
	\mathcal{T}^{(2)}_{\mu\nu}(x) \,*\, G\Big]\,.
	\raisetag{-1ex}
	\label{eq:integratedflowEMTstar}
\end{align}
In \cref{eq:integratedflowEMTstar} we have suppressed the dependence on the background $\bar{\phi}$. By $\mathcal{T}_{\mu\nu,\mathrm{init}}(x)$, we denote the EMT at the initial time $\tau=t_0$. As the causal temporal flow does not introduce a quantum violation of scale invariance, the respective anomalous terms cancel in the difference.  

For the solution of \cref{eq:integratedflowEMTstar} we note that structurally it is very similar to a  Dyson-Schwinger equation, for recent reviews see e.g.~\cite{Fischer:2018sdj, Huber:2018ned}. In particular, \cref{eq:integratedflowEMTstar} is a functional integral equation. Accordingly, it can be solved using the iteration techniques as used for DSEs: on the right hand side of \eq{eq:integratedflowEMTstar} we insert an appropriate guess for the solution, for instance the initial condition $\mathcal{T}_{\mu\nu,\mathrm{init}}(x)$. Then, we iterate \cref{eq:integratedflowEMTstar} until convergence. For this procedure it is useful to express the EMT and its derivatives, $\mathcal{T}^{(n)}$, in terms of the initial condition $\mathcal{T}^{(n)}_{\mathrm{init}}$ and the deviation $\Delta \mathcal{T}^{(n)}_\tau$. For instance, the second derivative reads  
\begin{align}
	\mathcal{T}^{(2)}_{\mu\nu,\tau}(x,a,b)
	\,=\,
	\mathcal{T}^{(2)}_{\mu\nu,\mathrm{init}}(x,a,b)
	\,+\,
	\Delta \mathcal{T}^{(2)}_{\mu\nu,\tau}(x,a,b) \,,
\end{align}
for general $\tau$. It has been shown in \cite{Corell:2019jxh, CorellDiss, HellerDiss}, that the causal constraints of the temporal flow allow us to classify the non-vanishing contributions to the flow according to the space-time structure of the interactions. Moreover, generically the local contributions dominate the flow.  Accordingly, we define 
\begin{align}\nonumber
	\Delta \mathcal{T}^{(2)}_{\mu\nu,\tau}(x,a,b) 
	=&\,\Delta \mathcal{T}^{(2)}_{\mu\nu,\mathrm{local}\,\tau}(x,a,b)\\[1ex]
	&\hspace{-2.4cm}+\Delta \mathcal{T}^{(2)}_{\mu\nu,\mathrm{nl}}(x,a,b)\, 
	\theta(\tau-x^0)\theta(\tau-a^0)\theta(\tau-b^0)\,.
	\label{eq:localEMT}
\end{align}
Here, all contributions containing $\deltac$\nobreakdash-functions are contained in $\Delta \mathcal{T}^{(2)}_{\mu\nu,\mathrm{local}\,\tau}$. Inserting \cref{eq:localEMT} in  \cref{eq:integratedflowEMTstar}, we notice that $\Delta \mathcal{T}^{(2)}_{\mu\nu,\mathrm{nl}}$ in the second line of \cref{eq:localEMT} does not satisfy the causal constraint of the EMT, \cref{eq:causalconstraintEMT}. Hence it does not contribute to the integrated flow, and the only contribution stems from the local term in \cref{eq:localEMT}.

\section{Integrated Flow and diagrammatic representations of the EMT}
\label{sec:consistency}

\begin{figure*}[t]
		\includegraphics[width=.7\textwidth]{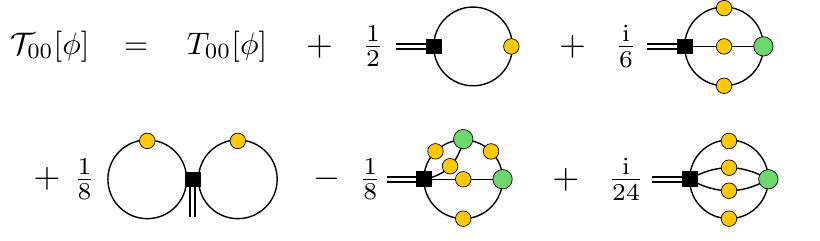}
		\caption{Diagrammatic representation of the EMT, \cref{eq:hamdiags}, using the diagrammatic depiction of propagators and vertices in \Cref{fig:hamdiags}, and dropping any reference to the renormalisation. We refer to the diagram involving $G^3$ as the \textit{sunset}. The diagram involving $G^2$ is called the \textit{eight}. The diagram involving $G^4$ and $\Gamma^{(3)}$ is called the \textit{squint-sunset}. The diagram with $G^4$ and $\Gamma^{(4)}$ is called the \textit{basketball}.}
		\label{fig:hamdiags}
\end{figure*}
The integrated flow \cref{eq:integratedflowEMTstar} constitutes another important result of the present work and offers a surprisingly simple diagrammatic form of the full energy-momentum tensor of the quantum field theory at hand. In the present Section we show how it can be used to recover the standard diagrammatic representation of the EMT in terms of its DSE. This derivation also shows the power of the analytical one-loop representation with the $*$-product. 

For this purpose we focus on the energy $\mathcal{T}_{00}$. For the present real scalar theory with the action \cref{eq:action},  the $00$-component of the classical EMT is read-off from \cref{eq:EMTscalar},  
\begin{widetext}
\begin{align}
	T_{00}[\varphi](x)
	=\frac{1}{2}\Big[ \big( \partial_{x^0}\varphi \big)^2 
	+ \big( \partial_{\vec{x}}\varphi \big)^2 + m^2\varphi^2\Big] + \frac{\lambda_3}{3!}\varphi^3 + \frac{\lambda_4}{4!}\varphi^4\,.
	\label{eq:classicalenergy}
\end{align}
The expectation values of powers of fields can be rewritten in terms of propagators and higher order correlation functions with \cref{eq:expval}. This leads us to 
\begin{align}\nonumber
	\mathcal{T}_{00}[\phi](x)
	=&T^\textrm{(ren)}_{00}[\phi](x) +\Biggl[ \frac{1}{2}\,T_{00}^{(2)}[\phi](x)\cdot G	
	+\frac{\imag}{6} \,T_{00}^{(3)}[\phi](x)\cdot G^3\cdot\Gamma^{(3)}\\[1ex]
	&+\frac{1}{8}\, T_{00}^{(4)}[\phi](x)\cdot G^2
	-\frac{1}{8}\, T_{00}^{(4)}[\phi](x)\cdot G^4 \cdot \Gamma^{(3)}\cdot G\cdot\Gamma^{(3)}
	+\frac{\imag}{24}\, T_{00}^{(4)}[\phi](x)\cdot G^4\cdot \Gamma^{(4)}\Biggr]_{\textrm{ren}}\,, 
	\label{eq:hamdiags}
 \end{align}
\end{widetext}
with $\phi$-dependent vertices $T_{00}^{(n)}[\phi](x)\defeq\delta T_{00}[\phi](x)/\delta\phi^{n}$. These vertices can be expanded in powers of the fields. For example, the second term in \cref{eq:hamdiags} reads
\begin{align}\nonumber
\frac{1}{2}\,T_{00}^{(2)}[\phi](x)\cdot G= & \frac12 \lim_{y \to x} \partial_{x} \partial_{y} G_{xy} 
+\frac12 m^2 G_{xx}\\[1ex]	
 &+\frac{1}{2}\, \lambda_3\phi_x G_{xx}
	+\frac{1}{4}\, \lambda_4\phi^2_x G_{xx}\,,
	\label{eq:hamphideriv}
\end{align}
where we used $\langle(\partial_{x}\varphi_x)(\partial_{x} \varphi_x) \rangle= \lim_{y\rightarrow x}\partial_{x}\partial_{y}\langle\varphi_x\varphi_y\rangle$. Moreover, the third term in the first line of \cref{eq:hamdiags} is proportional to $T_{00}^{(3)}[\phi]= \lambda_3 + \lambda_4 \phi$. 

We note that \cref{eq:hamdiags} is a formal expression as the diagrams require renormalisation, depending on the dimension $D$ of the theory. Moreover, part of the derivative terms in $T_{\mu\nu}$ come from the Lagrangian part proportional to $\eta_{\mu\nu}$ and renormalise accordingly. In turn, the other part comes from the composite operator $\partial_{\mu}\phi\,\partial_\nu\phi$ with its own renormalisation. It is indeed this difference which accounts for the trace anomaly and hence the breaking of quantum scale invariance. In the present work we are interested in the structural aspects of temporal flows, the renormalisation intricacies are discussed elsewhere in \cite{CausalFlowRen}. We indicate the respective anomalous terms with the superscript (ren) in $T^\textrm{(ren)}_{00}$ and the renormalisation of the diagrams with the respective subscript of the square bracket. In particular, summing over all diagonal components we have $T^\textrm{(ren)}{}_\mu{}^\mu=T_\mu{}^\mu+{\cal A}$ with the trace anomaly ${\cal A}$. 

Similar derivations also hold for all components $T_{\mu\nu}$ of the EMT, which leaves us with a closed three-loop exact expression for the quantum EMT. \Cref{eq:hamdiags} is depicted graphically in \Cref{fig:hamdiags}, using \Cref{fig:legend}. Its different graphs constitute different topologies in a diagrammatic expansion: Apart from the one-loop contribution we have a diagram with $G^2$ (\textit{eight}), a diagram with $G^3$ (\textit{sunset}), a diagram with $G^4$ and $\Gamma^{(3)}$ (\textit{squint-sunset}), and a diagram with $G^4$ and $\Gamma^{(4)}$ (\textit{basketball}). 

The integrated flow has to reproduce these different topologies as well as the respective prefactors. For identifying the relevant contributions of \cref{eq:hamdiags} to $\mathcal{T}^{(2)}_{00}$, we use the integrated flow \cref{eq:integratedflowEMTstar} and the causal constraint \cref{eq:causalconstraintEMT}: only terms that generate at least one $\deltac$ can contribute to the integrated flow. Thus, none of the field-derivatives of the terms in the second line in \cref{eq:hamdiags} have to be considered (the second line in \Cref{fig:hamdiags}). From the \textit{sunset}, only the contribution involving the explicit field dependence gives rise to a $\deltac$. Thus, the relevant local contributions to $\mathcal{T}^{(2)}_{00}$~are contained in
\begin{widetext}
\begin{align}
	\mathcal{T}^{(2)}_{00,\mathrm{local}}(x,a,b)=
	T^{(2)}_{00}(x,a,b) + \Delta \mathcal{T}^{(2)}_{00,\mathrm{local}}(x,a,b)\,, 
	\label{eq:T2local}
	\raisetag{-1ex}
\end{align}
where we again have dropped the reference to the necessary renormalisation. The $\phi$-derivatives of the kinetic term for $G$ are also non-local while the other two terms in \cref{eq:hamphideriv} generate local contributions. In summary, we obtain
	\begin{align}\nonumber
	\Delta \mathcal{T}^{(2)}_{00,\mathrm{local}}(x,a,b)
	=&\, 
	\frac{\imag}{2}\left[\lambda_3
	G^2_{x}\Gamma^{(3)}_a\sspace\deltac_{\!,xb} 
	+ (a\leftrightarrow b)\right]
	+\frac{\imag}{2}\left[\lambda_4
	\phi_x\, G^2_x\Gamma^{(3)}_a\sspace\deltac_{\!,xb} 
	+ (a\leftrightarrow b)\right]\\[1ex]
	&\,+\frac{1}{2}\, \lambda_4 G_{xx}\deltac_{\!xa}\deltac_{\!xb}
	+\frac{\imag}{6}\left[\lambda_4 G^3_x\Gamma^{(4)}_a\sspace\deltac_{\!,xb} 
	+\,(a\leftrightarrow b)\right]
	-\frac{1}{2}\left[\lambda_4 G^3_x\Gamma^{(3)}_a G\Gamma^{(3)}\sspace\deltac_{\!,xb} 
	+\,(a\leftrightarrow b)\right]\,.
	\label{eq:deltaT2}
\end{align}

For a concise representation we have used a short hand notation $[G^m_x F]$, where the dependence on any internal space-time indices has been dropped, see \Cref{app:[]}. The first term in \cref{eq:deltaT2} is generated from $\phi_x G_{xx}$. The second and the third term are generated from $\phi_x^2 G_{xx}$. The last two terms are generated by the relevant part of the \textit{sunset}. For all these terms, at least one field must be hit by the $\phi$\nobreakdash-derivative to generate a local contribution.
\end{widetext}
Now we insert \cref{eq:T2local} into the integrated flow \cref{eq:integratedflowEMTstar}. The arguments $a,b$ of all terms in \cref{eq:T2local} get closed by the $\tau$\nobreakdash-derivative of the propagator, $\partial_\tau G_{\tau,ab}$. In terms of loop topology we observe that closing the first two terms in \cref{eq:deltaT2} with $G_{ab}$ gives rise to the \textit{sunset}. The remaining terms generate the \textit{eight}, the \textit{basketball} and the \textit{squint-sunset}. In conclusion all topologies are present. 

{\renewcommand{\arraystretch}{2}
	\setlength{\arrayrulewidth}{.2mm}
	\begin{table}[t]
		\bigskip
		\centering 
		\begin{tabular}{c c c}
			diagram 	& $G\,*\,\Delta \mathcal{T}^{(2)}_{00,\mathrm{local}}$ 	&  factor \\[1pt] \hline
			\textit{eight} & $\partial_\tau G_{\tau,ab}\;G_{\tau,xx}\deltac_{\!,xa}\deltac_{\!,xb}$			&$\nicefrac{2}{4}$  \\
			\textit{sunset} & $\big[\sspace\partial_\tau G_{\tau,ab}\;G^2_{\tau,x}\deltac_{\!,x(a\leftrightarrow b)}\Gamma^{(3)}_{\tau,a}\sspace\big]$	&$\nicefrac{2}{3}$ \\  
			\textit{basketball}&$\big[\sspace\partial_\tau G_{\tau,ab}\;G^3_{\tau,x}\deltac_{\!,x(a\leftrightarrow b)}\Gamma^{(4)}_{\tau,a}\sspace\big]$&$\nicefrac{2}{4}$\\
			\textit{squint-sunset}&$\big[\sspace\partial_\tau G_{\tau,ab}\;G^3_{\tau,x}\deltac_{\!,x(a\leftrightarrow b)}\Gamma^{(3)}_{\tau,a} G_\tau\Gamma^{(3)}_\tau\sspace\big]$&$\nicefrac{2}{4}$
		\end{tabular}
		\smallskip
		\caption{
			Factors due to the evaluation of the $*$-product for the respective diagrams.
			The $*$-product was defined in 
			\cref{eq:Salmn}.
			Here, we use the following shorthand notation: $[\dots\deltac_{\!,x(a\leftrightarrow b)}\dots]=[\dots\deltac_{\!,xb}\dots]+[\dots\deltac_{\!,xa}\dots]$, cf. \cref{eq:deltaT2}.}
		\label{tab:starprod}
		\smallskip
	\end{table}
} 
It remains to compute the combinatoric factors. To that end we first note that the $*$-product in $T^{(2)}_{00}[\phibar](x)*G_{ab}[\phibar]$ reduces to the standard one,  $T^{(2)}_{00}[\phibar](x)\cdot G$, since $\phibar_x$ does not give rise to any $\theta_{\tau x}$. For the remaining diagrams of $\Delta \mathcal{T}^{(2)}_{00,\mathrm{local}}$, the $*$\nobreakdash-product has to be evaluated. The~corresponding contributions from the $*$\nobreakdash-product can be found in \Cref{tab:starprod}, and are derived by using \App{app:starproduct}. Combining the factors of \Cref{tab:starprod} with the ones already present in \cref{eq:deltaT2} and the overall factor of $\nicefrac{1}{2}$ of the flow, cf. \cref{eq:integratedflowEMTstar}, the integrated flow of the EMT reads
\begin{widetext}
	\begin{align}\nonumber
	\mathcal{T}_{00}[\phibar](x) -\mathcal{T}_{00,\mathrm{init}}[\phibar](x)
	=& \frac{1}{2}\,T_{00}^{(2)}[\phi](x)\cdot G	
	+\frac{\imag}{6} \,T_{00}^{(3)}[\phi](x)\cdot G^3\cdot\Gamma^{(3)}\\[1ex]
	&+\frac{1}{8}\, T_{00}^{(4)}[\phi](x)\cdot G^2
	-\frac{1}{8}\, T_{00}^{(4)}[\phi](x)\cdot G^4 \cdot \Gamma^{(3)}\cdot G\cdot\Gamma^{(3)}
	+\frac{\imag}{24}\, T_{00}^{(4)}[\phi](x)\cdot G^4\cdot \Gamma^{(4)}\,.
	\label{eq:integratedflowEMTdiags}
\end{align}
\end{widetext}
Thus, we already reproduced all diagrams of \cref{eq:hamdiags} with their correct prefactors. Note also that the difference on the left-hand side eliminates the anomalous contributions that sum up to the trace anomaly. The latter is not time-dependent and hence does not change in the temporal flow. This implies, that the right-hand side of \cref{eq:hamdiags} also does not encode the anomalous terms. What is left is to determine  $\mathcal{T}_{00,\mathrm{init}}[\phibar](x)=\mathcal{T}_{00,\tau=t_0}[\phibar](x)$, where we restrict ourselves to times $x^0> t_0$. For these times the initial $\mathcal{T}_{00,\mathrm{init}}$ does not comprise any fluctuations due to the causal regulator and it reduces to $T^{(\textrm{ren})}_{00}[\phibar](x)$ including the trace anomaly. In summary, we fully recover the diagrammatic representation \cref{eq:hamdiags} from the integrated flow. Moreover, \cref{eq:integratedflowEMTdiags} readily extends to all components of the EMT. Finally, as for the EMT-flow \cref{eq:causalflowEMTFinal}, the derivations above straightforwardly extend to $n$-point correlations $\langle T_{\mu_1\nu_1}\cdots T_{\mu_n\nu_n}\rangle$ of the EMT. Specifically, components of the two-point correlation functions are chiefly important for the derivation of transport coefficients.

\Cref{eq:integratedflowEMTdiags} is a non-trivial demonstration of the consistency of the t-fRG formalism as well as the computational simplicity of its unique one-loop exact representation: The integrated one-loop exact causal temporal flow of the EMT encodes the full diagrammatic representation \cref{eq:hamdiags} of the expectation value of the EMT. Moreover, this section also demonstrates the power of the causal constraints inherent to the t-fRG. These constraints greatly simplified the above derivation as the number of relevant diagrammatic contributions was reduced dramatically. Importantly, the $*$\nobreakdash-products, that encode the full causal structure of the flows, have a practical usability within numerical applications.

\section{Conclusion}
\label{sec:conclusion}

We have derived the flow of the energy-momentum tensor for general regulators. This equation contains terms that describe the explicit breaking of scale invariance by general regulators. We have shown, that for the causal temporal regulator these terms are absent since this regulator simply ensures a causal time evolution, and does not trigger any violation of scale invariance. Consequently, a potential trace anomaly of the EMT is described by providing renormalised initial conditions for the causal temporal flow. Alternatively one can augment the temporal regulator with a standard spatial infrared momentum regulator. Then the trace anomaly is generated by the spatial momentum flow. 

We have integrated the causal flow of the energy-momentum tensor analytically, which leads us to a one-loop excact functional relation for the energy-momentum tensor. With the help of a product of singular distributions put forward in \cite{Corell:2019jxh} this relation can be also implemented numerically. We have also demonstrated explicitly that it is consistent with the usual Dyson-Schwinger equation of the expectation value of the energy-momentum tensor.

\acknowledgments

We thank Lukas Corell and Nicolas Wink for discussions and collaborations on related subjects. MH thanks the HGSFP for financial support. This work is supported by EMMI, the BMBF grant 05P18VHFCA, and is part of and supported by the DFG Collaborative Research Centre SFB 1225 (ISOQUANT) as well as by the DFG under Germany's Excellence Strategy EXC - 2181/1 - 390900948 (the Heidelberg Excellence Cluster STRUCTURES).

\appendix

\section{Useful Identities for Metric-Derivatives}
\label{app:identities}
 
Here, we display some useful identities involving the metric-derivative.
For the derivative of the metric and its determinant, we use the standard identities

\begin{align}\label{eq:metricderivative}
	\frac{\delta g^{\alpha\beta}(a)}{\delta g^{\mu\nu}(b)}
	&=\Big( \delta^\alpha_\mu\delta^\beta_\nu + \delta^\alpha_\nu\delta^\beta_\mu \Big)\deltac(a-b)\\[1em]
	\frac{\delta\sqrt{-g(a)}}{g^{\mu\nu}(b)}
	&=\,-\frac{1}{2}\,\sqrt{-g(a)}\,
	g_{\mu\nu}(a)\deltac(a-b)\,.
	\label{eq:detderivative}
\end{align}
Using these two relations, it is straightforward to derive
\begin{align}
	\frac{\delta }{\delta g^{\mu\nu}(b)}
	\left[\int_{\contourC,a}\!\!\!\!\!\sqrt{-g(a)}\right]=-\frac{1}{2} \int_{\contourC,a}\!\!\!\!\!\sqrt{-g(a)}\, g_{\mu\nu}(a)\deltac(a-b)\,.
	\label{eq:intderivative}
\end{align}
The manifestly covariant $\delta$-function is defined by
\begin{align}
	\oneO_{ab}= \frac{1}{\sqrt{-g(a)}}\deltac(a-b)
	\label{eq:one}
\end{align}
Accordingly, its metric-derivative reads
\begin{align}\nonumber
	\frac{\delta\,\oneO_{ab} }{\delta g^{\mu\nu}(x)}
	&= \frac{\delta }{\delta g^{\mu\nu}(x)}
	\left(\frac{\deltac(a-b)}{\sqrt{-g(a)}}\right)\\[1em]
	&= \, \frac{1}{2}\,g_{\mu\nu}(a)
	\deltac(a-x)\,\oneO_{ab}\;.
	\label{eq:onederivative}
\end{align}
These relations ensure that 
\begin{align}
	\frac{\delta }{\delta g^{\mu\nu}(x)}
	\bigg(
	\int_{\contourC}\sqrt{-g} \;\oneO
	\bigg)= 0 \,.
\end{align}

In \Cref{sec:causalflow}, we need the metric-derivative of the regulator $R_\tau$ and that of $\partial_{\tau} R_\tau$. Both are proportional to $\oneO$ and have no further metric dependence. Hence, we find with  \cref{eq:onederivative}, 
\begin{align}
	\frac{1}{\sqrt{-g(x)}}\frac{\delta (R_\tau\,,\, \partial_{\tau} R_\tau)_{ab}}{\delta g^{\mu\nu}(x)}
	=\frac{1}{2}g_{\mu\nu}(x)\,(R_\tau\,,\, \partial_{\tau} R_\tau)_{ab} \oneO_{ax} \,, 
	\label{eq:metricderivativereg}
\end{align}
where no sum over the index $a$ is implied. 

\section{The $*$-product}
\label{app:starproduct}

The $*$\nobreakdash-product accounts for the fact that the regulator, the propagator and the other correlators can have coincident singularities \cite{Corell:2019jxh, HellerDiss}, leading to contributions of the form $\delta(\tau-x^0)\,*\,\theta(\tau-x^0)$. These singularities have to be treated with care. To that end, we introduce a regularisation $\varepsilon$ for the distributions present. Then, such products lead to, e.g.\ \cite{Corell:2019jxh},
\begin{subequations} \label{eq:Salmn}
\begin{align}
	&\delta_{\tau x} \,*\, \theta_{\tau x}^n\;\defeq\;
	\lim_{\varepsilon\to 0}\,\delta_{\varepsilon,\tau x}\, [\theta_{\varepsilon,\tau x}]^n\,,
\end{align}
with
\begin{align}
\lim_{\varepsilon\to 0}\,\delta_{\varepsilon,\tau x}\, [\theta_{\varepsilon,\tau x}]^n=\delta_{\tau x}\,	\int_0^1 dx\, x^n = \frac{1}{n+1}\; \delta_{\tau x} \,.
\end{align}
\end{subequations}
Here, $\theta_{\varepsilon,\tau x}\defeq\theta_\varepsilon(\tau-x^0)$ and similarly for the $\delta$\-/distribution. Crucially, the $\delta$- and $\theta$-functions in \cref{eq:Salmn} share the same regularisation $\varepsilon$ as distributions. This is due to the fact that this regularisation is inherited from the regularisation of the sharp cutoff, $R_\tau\rightarrow R_{\varepsilon,\tau}$. Thus, the product of  distributions that come from different $n$-point functions is uniquely defined in terms of \cref{eq:Salmn}. We remark that identities like \cref{eq:Salmn} have also been used in momentum space fRGs, see e.g.\ \cite{Morris:1993qb, Metzner:2011cw}.

\section{Expectation Values}
\label{app:diags}

For the computation of the expectation value of the classical energy \cref{eq:classicalenergy}, cf.~\cref{eq:hamdiags} and \Cref{fig:hamdiags} respectively, we 
use the functional relation (see e.g.~\cite{Pawlowski:2005xe}),
\begin{align}
	\left\langle \prod_{i=1}^n \varphi(x_i)\right\rangle =  \prod_{i=1}^n\left[
	\int\displaylimits_{\mathrlap{\contourC,z_i}} 
	 G[\phi](x_i , z_i)\frac{\delta }{\delta \phi(z_i) }+
	\phi(x_i)\right] \,.
	\label{eq:expval}
\end{align}
%

\section{Shorthand Notation for Loops}
\label{app:[]}
Consider a product of $m$ propagators in a diagram, given by $G_{x1}\cdots G_{xm}$ and fully contracted with another tensor $ F_{1\dots m}$, to wit, 
\begin{align}
	[G^m_x F]=G_{x1}\cdots G_{xm}F_{1\dots m}\,,
\end{align}
and $F_{1\dots m}$ comprises the rest of the diagram. For instance,
\begin{align}
	[G^2_{x}\Gamma^{(3)}_a\sspace\deltac_{\!,xb} ]=G_{xz_1}G_{xz_2}\,\Gamma^{(3)}_{az_1z_2}\sspace\deltac_{\!,xb} \,,
	\label{eq:diagbracket}
\end{align}
where sums or integrals over the indices $z_1,z_2$ are implied. 

\vfill

\bibliography{flow_EMT}

\end{document}